\documentclass[conference]{IEEEtran}
\ifCLASSINFOpdf
\else
\fi
\usepackage{enumitem}
\setlist[description]{font=\normalfont\itshape\space}

\usepackage{pgfplots}
\usepgfplotslibrary{external,colormaps,groupplots,statistics}
\pgfplotsset{compat=1.8}
\usepackage{tikz}
\usepackage{listings}
\usetikzlibrary{arrows,shapes,positioning}
\usepackage{multirow}
\usepackage{graphicx}
\usepackage{microtype}
\usepackage{balance}
\usepackage{xcolor}
\usepackage{xspace}
\usepackage{booktabs}

\definecolor{codegreen}{rgb}{0.25,0.5,0.35}
\definecolor{codegray}{rgb}{0.5,0.5,0.5}
\definecolor{codepurple}{rgb}{0.6,0,0}
\definecolor{backcolour}{rgb}{0.95,0.95,0.92}
\definecolor{colorstring}{rgb}{0.5,0,0.35}	
\definecolor{rltred}{rgb}{0.5,0,0}
\definecolor{rltgreen}{rgb}{0,0.5,0}
\definecolor{rltblue}{rgb}{0,0,0.5}
\definecolor{DarkGreen}{rgb}{0.00,0.60,0.00}
\definecolor{ScarletRed}{rgb}{0.80,0.00,0.00}
\definecolor{blizzardblue}{rgb}{0.67, 0.9, 0.93}
\definecolor{green-yellow}{rgb}{0.68, 1.0, 0.18}
\definecolor{dkgreen}{rgb}{0,0.6,0}
\definecolor{gray}{rgb}{0.5,0.5,0.5}
\definecolor{mauve}{rgb}{0.58,0,0.82}
\definecolor{lightgrey}{rgb}{0.90,0.90,0.90}
\definecolor{grey}{gray}{0.75}
\definecolor{light-gray}{gray}{0.80}

\lstdefinestyle{mystyle}{
	backgroundcolor=\color{backcolour},   
	commentstyle=\color{codegreen},
	keywordstyle=\color{colorstring}\bfseries,
	numberstyle=\ttfamily\color{codegray},
	stringstyle=\color{codepurple},
            basicstyle={\scriptsize\ttfamily},
	breakatwhitespace=false,         
	breaklines=true,                 
	captionpos=b,                    
	keepspaces=true,                 
	numbers=none,                    
	showspaces=false,                
	showstringspaces=false,
	showtabs=false,                  
	tabsize=2
}
\usepackage{boxedminipage}
{\smallskip
\noindent
\let\emph=\textbf
\begin{boxedminipage}{\columnwidth}\begin{center}\em}%
{\end{center}\end{boxedminipage}%
}

\usepackage{microtype}
\DisableLigatures{}

\newcommand{\evo}{{\sc EvoMaster}\xspace}

\begin{document}

\title{EvoMaster: Evolutionary Multi-context Automated System Test Generation}

\author{
\IEEEauthorblockN{Andrea Arcuri}
\IEEEauthorblockA{Westerdals Oslo ACT, Oslo, Norway\\
and SnT, University of Luxembourg, Luxembourg\\
Email: arcand@westerdals.no}
}


%Telephone: (800) 555--1212, Fax: (888) 555--1212}


\maketitle

\begin{abstract}
This paper presents \evo, an open-source tool that is able to automatically generate
system level test cases using evolutionary algorithms.
Currently, \evo targets RESTful web services running on JVM technology,
and has been used to find several faults in existing open-source projects.
We discuss some of the  architectural decisions made for its implementation,
 and future work.

\end{abstract}

% \fi
\IEEEpeerreviewmaketitle

%%%%%%%%%%%%%%%%%%%%%%%%%%%%%%%%%%%%%%%%%%%%%%%%%%%%%%%%%%%%%%%%%%%%%%%%%%%%%%%%%%%%%%%%%%
\section{Introduction}

There exist different tools that can automatically generate unit tests, using variants of random testing (e.g., Randoop~\cite{PLEB07}), evolutionary search (e.g., EvoSuite~\cite{fraser2011evosuite}) or dynamic symbolic execution (e.g., Pex/IntelliTest~\cite{TiH08}).
For smartphone applications~\cite{android2015}, there are tools like Sapienz~\cite{mao2016sapienz} that can generate sequences of events on the GUI.
For web applications serving HTML pages, there are web crawler tools like Crawljax~\cite{mesbah2012invariant}. 
These crawlers can be used for testing of web applications, but they are black-box, and do not take into account the internal details of the server side code.   
Furthermore, little exists that is \emph{available} (i.e., a tool that can be downloaded and used) for \emph{white-box} system testing of enterprise applications~\cite{arcuri2018experience}, in particular RESTful web services.

This paper introduces \evo, a new tool that aims at test generation at system level using evolutionary techniques, in particular the MIO algorithm~\cite{mio2017}.
At the current stage, \evo targets RESTful APIs~\cite{fielding2000architectural,allamaraju2010restful} running on JVMs.
However, \evo is architectured in a way in which it can be extended for other languages and other system test contexts.

\begin{figure}[t]
  \centering
  \includegraphics[width=.45\textwidth]{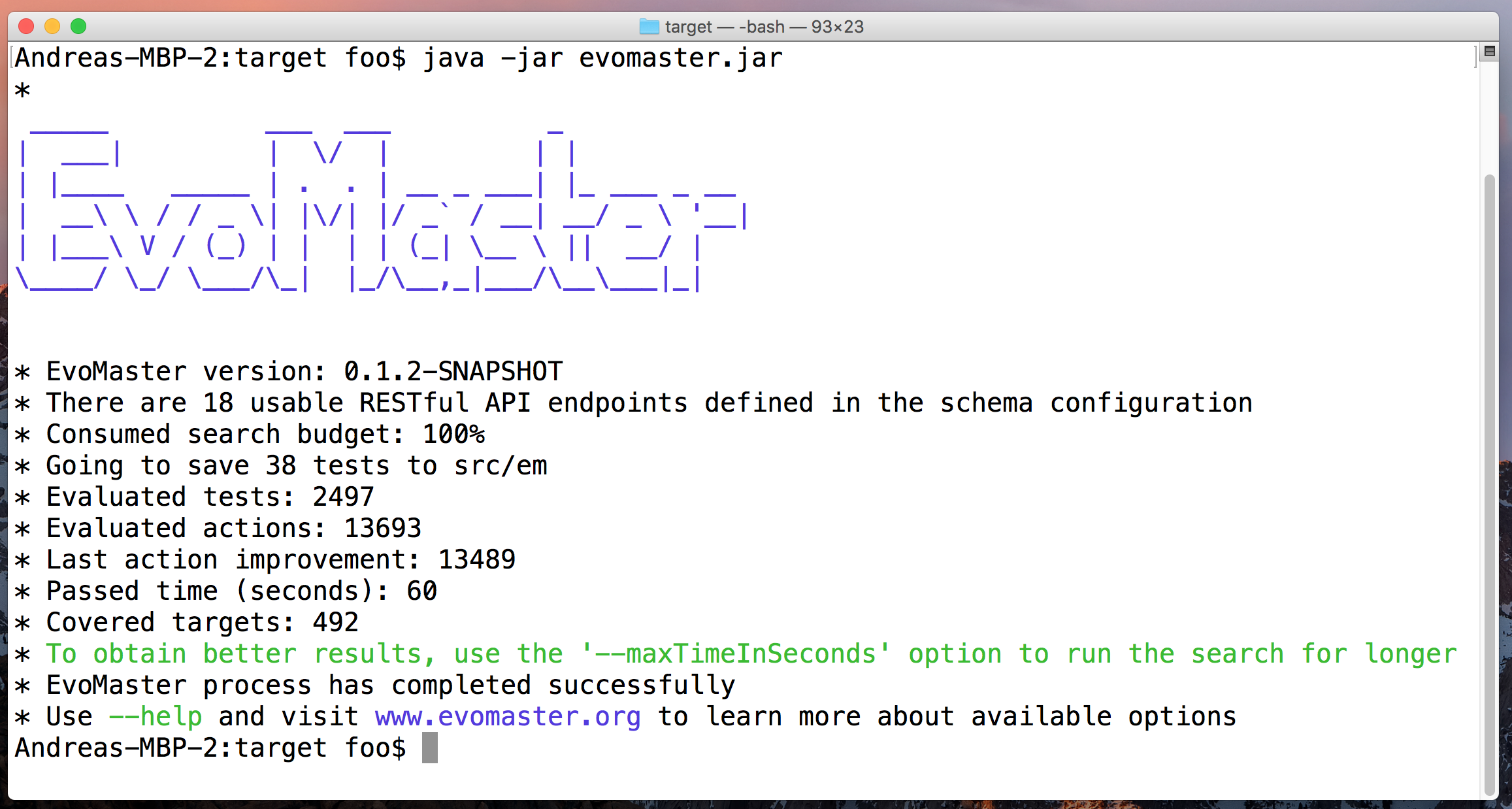}
  \caption{\label{fig:terminal}
  	Usage of \evo from command terminal.
  }
\end{figure}

\begin{figure}[t]
\begin{lstlisting}[language=java,basicstyle=\tiny]
@Test
public void test12() throws Exception {
        
    String location_activities = "";
        
    String id_0 = given().accept("*/*")
            .header("Authorization", "ApiKey administrator") // administrator
            .contentType("application/json")
            .body("{\"name\":\"JIg\", 
                  \"date_updated\":\"1968-7-28T10:40:58.000Z\",
                  \"description_material\":\"CDasIs\", 
                  \"description_prepare\":\"VatRg\", 
                  \"description_main\":\"vbhUS\",       
                  \"description_safety\":\"mdMZXtHaW6Ac0L7\", 
                  \"age_min\":-1639552914, 
                  \"age_max\":-546, 
                  \"participants_min\":-166, 
                  \"time_max\":-728, 
                  \"featured\":true, 
                  \"activity\":{\"ratings_sum\":-2169794882535544017, 
                                \"favourites_count\":2018287764382358555, 
                                \"ratings_average\":0.7005221066369205, 
                                \"related\":[5230990194698818394, 
                                             4025421724722458028, 
                                             -1291838056, 
                                             -210322044]}}")
            .post(baseUrlOfSut + "/api/v1/activities")
            .then()
            .statusCode(200)
            .extract().body().path("id").toString();
                
    location_activities = "/api/v1/activities/" + id_0;
        
    given().accept("*/*")
            .header("Authorization", "ApiKey administrator") // administrator
            .contentType("application/json")
            .body("{\"rating\":7126434, \"favourite\":false}")
            .post(resolveLocation(location_activities, 
                  baseUrlOfSut + "/api/v1/activities/-324163273/rating"))
            .then()
            .statusCode(204);
        
    given().accept("*/*")
            .header("Authorization", "ApiKey administrator") // administrator
            .delete(resolveLocation(location_activities, 
                    baseUrlOfSut + "/api/v1/activities/-324163273/rating"))
            .then()
            .statusCode(204);
}
\end{lstlisting}
\caption{\label{fig:test}
Example of test generated by \evo.
}
\end{figure}

Modern  web applications often rely on external \emph{web services}.
Large and complex enterprise applications can be split into  individual web service components,
in what is typically called a \emph{microservice} architecture~\cite{newman2015building}.
The assumption is that individual components are easier to develop and maintain compared to a large monolithic application.
The use of microservice applications is a very common practice in industry, done for example in companies like
 Netfilx, Uber, Airbnb, eBay, Amazon, Twitter, Nike, etc~\cite{rajesh2016spring}.

Besides being used internally in many enterprise applications, there are many web services available on the Internet.
Websites like 
\emph{ProgrammableWeb}\footnote{https://www.programmableweb.com/api-research} 
currently list more than 16 thousand Web APIs.
Many companies provide APIs to their tools and services using REST, which is currently the most common type of web service,
like for example
Google\footnote{https://developers.google.com/drive/v2/reference/},
Amazon\footnote{http://docs.aws.amazon.com/AmazonS3/latest/API/Welcome.html},
Twitter\footnote{https://dev.twitter.com/rest/public},
Reddit\footnote{https://www.reddit.com/dev/api/},
LinkedIn\footnote{https://developer.linkedin.com/docs/rest-api},
etc.

Testing web services, and in particular RESTful web services, does pose many challenges~\cite{canfora2009service,bozkurt2013testing}.
Different techniques have been proposed.
However, most of the work so far in the literature has been concentrating on black-box testing of SOAP web services, and not REST~\cite{arcuri2017restful}.

Figure~\ref{fig:terminal} shows a use of \evo from command terminal, whereas 
Figure~\ref{fig:test} shows an example of generated test in Java using the highly popular RestAssured\footnote{https://github.com/rest-assured/rest-assured} library  (which helps in writing tests that require HTTP calls).
Automatically generating tests for RESTful APIs is a complex task, because a test might require several HTTP calls.
Each HTTP call might require to set up the right URL (path and query parameters), HTTP headers and an HTTP payload body.
This latter can be particularly complex, as the RESTful API could take as input any arbitrary kind of data (usually in JSON or XML format).
Furthermore, a HTTP call might require data from the output of a previous HTTP call.
This is a typical example when a resource is created on the server with a HTTP POST request, and then the returned id of this resource is needed to have a GET request on such newly generated resource. 
A tool aiming at generating this kind of tests needs to be able to handle all of these cases.

Although \evo is still in an early phase of development (it was started in the late 2016),
it has already been used to successfully find several bugs in existing open-source projects and in an industrial application~\cite{arcuri2017restful}.
\evo is released under the LGPL open-source license, and it is freely accessible on 
GitHub\footnote{https://github.com/EMResearch/EvoMaster}.

%This paper TODO

%The rest of this paper discusses: some of the tecnical implementation details of \evo(Section~\ref{sec:implementation}); manual prepartions needed before using it (Section~\ref{sec:manual}); some of its available configurations (Section~\ref{sec:config}); current results (Section~\ref{sec:results}).
%Finally, Section~\ref{sec:conclusion} concludes the paper.  

%-----------------------------------------------------------------------------------------
\section{Tool Implementation}
\label{sec:implementation}

\begin{figure}[t]
  \centering
  \includegraphics[width=.45\textwidth]{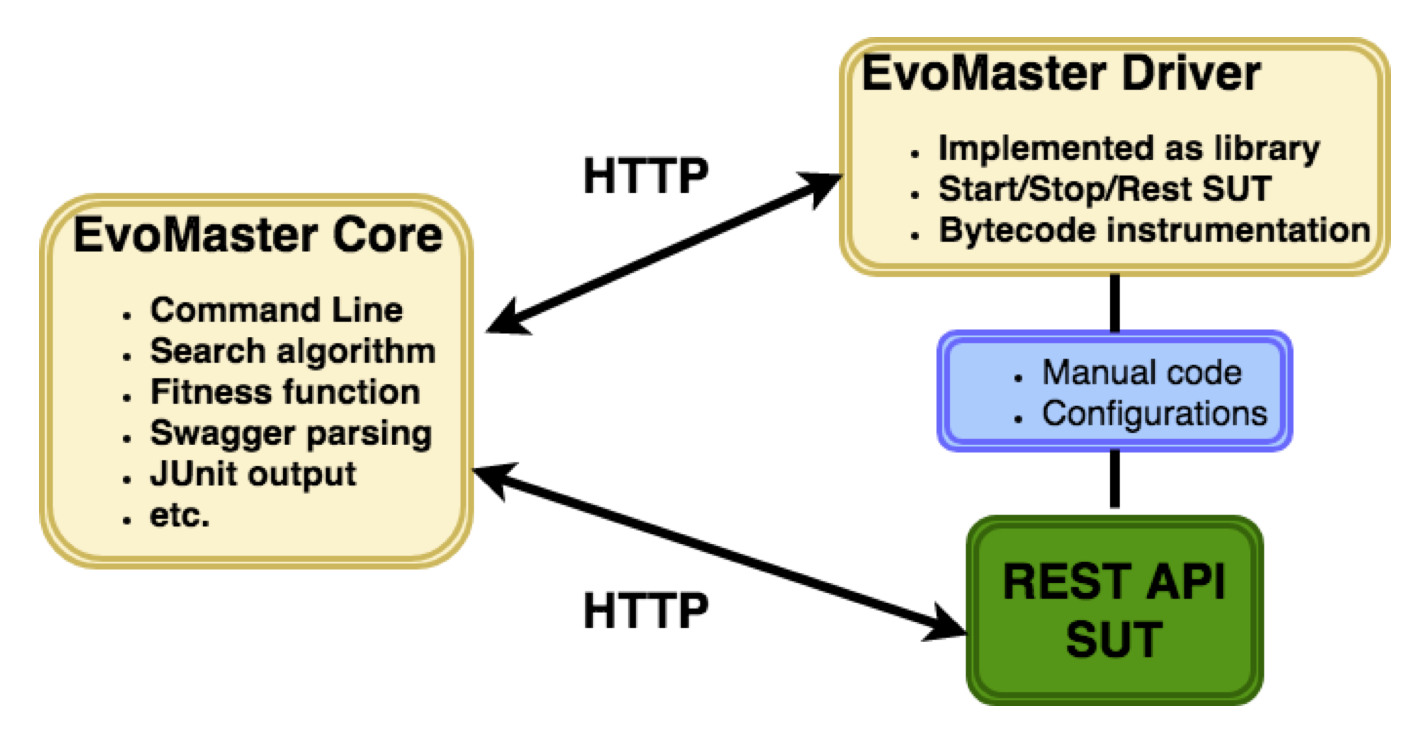}
  \caption{\label{fig:architecture}
	High level architecture of \evo.
  }
\end{figure}

\evo is composed of two main components: a \emph{core} process responsible for the main functionalities (e.g., command-line parsing, search and generation of test files), and a \emph{driver} process.
This latter is responsible to start/stop/reset the system under test (SUT) and instrument its source code, e.g., via automated bytecode manipulation, in a similar way of how unit test tools like EvoSuite~\cite{fraser2011evosuite} do.
For example, you need to add probes in the bytecode to check which statements are executed, and also to define heuristics to help solving the predicates in the branch statements (e.g., the so called \emph{branch distance}~\cite{Mcm04}).
Such test execution information is then exported by the \emph{driver} module (in JSON format) and used by
the \emph{core} process to generate new test cases.
Figure~\ref{fig:architecture} shows a high level overview of \evo's architecture.

\evo implements different kinds of search algorithms for test suite generation (e.g., WTS~\cite{GoA_TSE12} and MOSA~\cite{dynamosa2017}), where MIO~\cite{mio2017} is the default one.
\evo generates test suites with the goal of optimising white-box, code coverage metrics (e.g., statement and branch coverage) and fault detection (e.g., HTTP 5xx status codes can be used in some cases as automated oracles).
Each test will be composed of one or more HTTP calls.
The generated test files (e.g., using 
JUnit\footnote{http://junit.org/junit4/} 
and 
RestAssured\footnote{https://github.com/rest-assured/rest-assured} 
libraries)
are self-contained, as using the \evo driver as a library to automatically start the SUT before running the tests (e.g., in JUnit this can be done in a \texttt{@BeforeClass} init method).

The \emph{core} process of \evo is written in Kotlin, a new language that can compile into JVM bytecode.
The choice of Kotlin was due to the fact that we consider it as the best language for developing tools like \evo.
On the other hand, the drivers need to be implemented based on the target language of the SUT.
Currently, we provide a driver only for JVM languages (e.g., Java and Kotlin).
Adding support for a new language (e.g., C\#) does not require any change in the \emph{core} process, as communications between \emph{core} and \emph{driver} are program language agnostic (e.g., JSON over HTTP).

To use \evo on a given SUT, a test engineer has to provide in a configuration class some basic information, like for example where to find the SUT's executable (e.g., a uber jar) and on which TCP port the started RESTful service will listen on.
This is discussed in more details in the next section.

%-----------------------------------------------------------------------------------------
\section{Manual Preparations}
\label{sec:manual}

\begin{figure}[!t]
\caption{\label{fig:rest_controller}
Example of class that needs to be implemented by the developers of the SUT to enable the usage of our test case generation tool.
In this particular case, the SUT is written with SpringBoot, where \emph{Application} is the main entry point of the SUT.
}
\begin{lstlisting}[language=java,basicstyle=\tiny]
public class EMController extends EmbeddedSutController {

    private ConfigurableApplicationContext ctx;
    private final int port;
    private Connection connection;

    public static void main(String[] args) {

        EMController controller = new EMController(port);
        InstrumentedSutStarter starter = new InstrumentedSutStarter(controller);

        starter.start();
    }


    public EMController(){this(0);}

    public EMController(int port) {
        this.port = port;
    }

    @Override public int getControllerPort(){
        return port;
    }

    @Override public String startSut() {

        ctx = SpringApplication.run(Application.class,
                new String[]{"--server.port=0"});

        if(connection != null){
            try { connection.close();
            } catch (SQLException e) {
                e.printStackTrace();
            }
        }
        JdbcTemplate jdbc = ctx.getBean(
                                   JdbcTemplate.class);
        try {
            connection = jdbc.getDataSource()
                             .getConnection();
        } catch (SQLException e) {
            e.printStackTrace();
        }
        return "http://localhost:"+getSutPort();
    }

    protected int getSutPort(){
        return (Integer)((Map) ctx.getEnvironment()
                .getPropertySources().get("server.ports")
                .getSource()).get("local.server.port");
    }

    @Override public boolean isSutRunning() {
        return ctx!=null && ctx.isRunning();
    }

    @Override public void stopSut() { ctx.stop();}

    @Override public String getPackagePrefixesToCover() {
        return "org.javiermf.features.";
    }

    @Override public void resetStateOfSUT() {
        ScriptUtils.executeSqlScript(connection, 
            new ClassPathResource("/empty-db.sql"));
        ScriptUtils.executeSqlScript(connection, 
            new ClassPathResource("/data-test.sql"));
    }

    @Override public String getUrlOfSwaggerJSON() {
        return "http://localhost:"+getSutPort()+
           "/swagger.json";
    }

    @Override public List<AuthenticationDto> 
                        getInfoForAuthentication(){
        return null;
    }
}
\end{lstlisting}
\end{figure}

In contrast to tools for unit testing like EvoSuite, which are 100\% fully automated (a user just need to select in their IDE for which classes the tests should be generated),
our tool \evo for system/integration testing of RESTful APIs does require some manual configuration.
This is not a limitation of the tool, but rather one of challenges of system-level testing.

The developers of the RESTful APIs need to import our library (published on Maven Central Repository\footnote{https://mvnrepository.com/artifact/org.evomaster/evomaster-client-java}), 
and then create a class that extends the \emph{EmbeddedSutController} class in such library.
The developers will be responsible to define how the SUT should be started, where the Swagger schema can be found (which defines what present in the API), which packages should be instrumented, etc.
This will of course vary based on how the RESTful API is implemented, e.g.,
if with 
Spring\footnote{https://github.com/spring-projects/spring-framework}, 
DropWizard\footnote{https://github.com/dropwizard/dropwizard}, 
Play\footnote{https://github.com/playframework/playframework}, 
Spark\footnote{https://github.com/perwendel/spark} 
or Java EE.

Figure~\ref{fig:rest_controller} shows an example of one such class we had to write for one of the SUTs in our empirical studies.
That SUT uses SpringBoot. 
That class is quite small, and needs to be written only once.
It does not need to be updated when there are changes internally in the API.
The code in the superclass \emph{EmbeddedSutController} will be responsible to do the automatic bytecode instrumentation of the SUT, and it will also start a RESTful service to enable our testing tool to remotely call the methods of such class.

However, besides starting/stopping the SUT and providing other information (e.g., location of the Swagger file), there are two further tasks the developers need to perform:
\begin{itemize}
\item	RESTful APIs are supposed to be stateless (so they can easily scale horizontally), but they can have side effects on external actors, such as a database.
In such cases, before each test execution, we need to reset the state of the SUT environment.
This needs to be implemented inside the \emph{resetStateOfSUT()} method.
In the particular case of the class in Figure~\ref{fig:rest_controller},
two SQL scripts are executed: one to empty the database, and one to fill it with some existing values.
We did not need to write those scripts by ourself, as we simply re-used the ones already available in the manually written tests in that SUT.
How to automatically generate such scripts would be an important topic for future investigations.
\item If a RESTful API requires some sort of authentication and authorization, such information has to be provided by the developers in the \emph{getInfoForAuthentication()} method.
For example, even if a testing tool would have full access to the database storing the passwords for each user, it would not be possible to reverse engineer those passwords from the stored hash values. 
Given a set of valid credentials, the testing tool will use them as any other variable in the test cases, e.g., to do HTTP calls with and without authentication.
\end{itemize}

Once such class is implemented, it needs to be run as a process (see its \emph{main} method).
This can be easily done in an IDE like IntelliJ/Eclipse by right-clicking on it.
Once this driver process is started, it will open a listening TCP port.
We can then start the \evo executable from a command terminal 
(e.g., recall Figure~\ref{fig:terminal}), which will connect to the driver process via TCP, and start generating test cases.
The documentation of \evo at \texttt{www.evomaster.org} provides links to videos on how to do these steps.

To enable researchers to use \evo in their experiments, we have provided on GitHub\footnote{https://github.com/EMResearch/EMB} 
a set of open-source projects for which we maintain the \evo driver classes needed to use it.
Note: as the \emph{driver} modules provide test execution information and heuristics independently from the \emph{core} process, such drivers can also be used in other system testing tools besides \evo.
This is of particular importance, as writing a bytecode manipulation library is a complex task.

Besides \emph{EmbeddedSutController}, users have also the option of rather extending the \emph{ExternalSutController} class.
This latter case is to handle situations in which it is not easy, or even possible, to start a web service directly from a class (e.g., Java EE).
To handle these cases, we enable the option to start the SUT on a separate, external process from the driver one, instead of running the SUT embedded in the same process of the driver.
To do so, we need the SUT to be packaged in a self-executable jar file.
The \evo driver library will \emph{automatically} handle all the necessary technical details on how to start/stop such process, enable JavaAgents, and collect statistics from these spawn processes.

%%%%%%%%%%%%%%%%%%%%%%%%%%%%%%%%%%%%%%%%%%%%%%%%%%%%%%%%%%%%%%%%%%%%%%%%%%%%%%%%%%%%%%%%%%
\section{Configurations}
\label{sec:config}

\evo has several configurations, which can be set with command line options.
For a practitioner, the main options are:

\begin{description}

\item[--help]: List all available options.

\item[--maxTimeInSeconds $<$Int$>$]:
Maximum number of seconds allowed for the search. The more time is allowed, the better results one can expect. But then the test generation will take longer.

\item[--outputFolder $<$String$>$]:
The path directory of where the generated test classes should be saved to.

\item[--outputFormat $<$OutputFormat$>$]:
Specify in which format the tests should be outputted. 
For example, JAVA\_JUNIT\_5 or JAVA\_JUNIT\_4.

\item[--testSuiteFileName $<$String$>$]:
The name of the generated file with the test cases.

\end{description}

All options provide sensible default values.
For example, by default the search lasts one minute.

For researchers, most the of internal settings of the search algorithms (e.g., population size) can be configured via command line options, like the different parameters used in the MIO algorithm~\cite{mio2017}.

%%%%%%%%%%%%%%%%%%%%%%%%%%%%%%%%%%%%%%%%%%%%%%%%%%%%%%%%%%%%%%%%%%%%%%%%%%%%%%%%%%%%%%%%%%
\section{Current Results}
\label{sec:results}

\evo was evaluated in~\cite{arcuri2017restful} on three different RESTful APIs: two open-source, and one from our industrial partners.
These APIs were between 2 and 10 thousand lines of Java code.

On such APIs, \evo found 38 unique bugs, where HTTP calls were generated in a way in which 5xx (server error, internal crash) HTTP codes were returned by the SUT responses.
However, on such SUTs the statement code coverage was only between 20\% and 40\%.
One main reason is that these SUTs (and RESTful APIs in general) interact with databases.
Supporting databases in search-based software testing (e.g., heuristics based on the results of the SQL queries) is one of current main activities in the \evo development.

%%%%%%%%%%%%%%%%%%%%%%%%%%%%%%%%%%%%%%%%%%%%%%%%%%%%%%%%%%%%%%%%%%%%%%%%%%%%%%%%%%%%%%%%%%
\section{Conclusion}
\label{sec:conclusion}

In this paper, we have presented \evo, a new tool that aims at generating white-box, system-level test cases for enterprise/web applications.
This type of systems are very common in industry.
But, in contrast to unit and mobile testing, to the best of our knowledge there is no available existing \emph{white-box} tool that addresses enterprise/web applications. 

Internally, \evo uses evolutionary techniques, like the MIO algorithm~\cite{mio2017}.
Currently, \evo does target RESTful APIs, but it is architectured in a way in which it will be easily extended to other contexts.
For example, the bytecode instrumentation is released as a library on Maven Central Repository, and can be integrated in other tools.

This paper describes some of the technical details of \evo, current results (e.g, bugs found in existing APIs) and future work (supporting SQL databases). 
To enable technology transfer from academic research to industrial practice, \evo is released with  a permissive open-source license (LGPL v3.0), and published on GitHub.
To learn more about \evo, visit our webpage at:
\texttt{www.evomaster.org}

%%%%%%%%%%%%%%%%%%%%%%%%%%%%%%%%%%%%%%%%%%%%%%%%%%%%%%%%%%%%%%%%%%%%%%%%%%%%%%%%%%%%%%%%%%
\section*{Acknowledgment}
This work is supported by the National Research Fund, Luxembourg (FNR/P10/03).



%\bibliographystyle{IEEEtran}

\bibliographystyle{IEEEtran}
\bibliography{../../papers}

\end{document}